\documentclass[manuscript,screen]{acmart}
\AtBeginDocument{%
  \providecommand\BibTeX{{%
        \normalfont B\kern-0.5em{\scshape i\kern-0.25em b}\kern-0.8em\TeX}}}

\setcopyright{rightsretained}
\copyrightyear{2023}
\acmYear{2023}
\acmDOI{}
\acmConference[CCAI 2023]{CHI 2023 Workshop on Child-centred AI Design: Definition, Operation and Considerations}{April 23, 2023}{Hamburg, Germany}
\acmBooktitle{CHI 2023 Workshop on Child-centred AI Design: Definition, Operation and Considerations, April 23, 2023, Hamburg, Germany}

\begin{document}

\title{Towards Building Child-Centered Machine Learning Pipelines: Use Cases from K-12 and Higher-Education}

\author{Alpay Sabuncuoğlu}
\email{asabuncuoglu13@ku.edu.tr}
\affiliation{%
  \institution{UNVEST R\&D Center}
  \city{Istanbul}
  \country{Turkey}
}

\author{Ceylan Beşevli}
\email{c.besevli@ucl.ac.uk}
\affiliation{%
  \institution{University College London (UCL)}
  \city{London}
  \country{United Kingdom}
}

\author{T. Metin Sezgin}
\email{mtsezgin@ku.edu.tr}
\affiliation{%
  \institution{Koc University}
  \city{Istanbul}
  \country{Turkey}
}

\begin{abstract}
    Researchers and policy-makers have started creating frameworks and guidelines for building machine-learning* (ML) pipelines with a human-centered lens \cite{microsoft_guideline, ieee_standard}. Machine Learning pipelines stand for all the necessary steps to develop ML systems (e.g., developing a predictive keyboard) \cite{case_studies}. On the other hand, a child-centered focus in developing ML systems has been recently gaining interest as children are becoming users of these products \cite{unesco_policy_2021}. These efforts dominantly focus on children’s interaction with ML-based systems. However, from our experience, ML pipelines are yet to be adapted using a child-centered lens. In this paper, we list the questions we ask ourselves in adapting human-centered ML pipelines to child-centered ones. We also summarize two case studies of building end-to-end ML pipelines for children’s products.

    \textit{\textbf{*}Throughout the paper, we used the term ML rather than AI. Commercial examples use the term AI loosely to define all the systems that learn from data. However, the term AI covers a broad range of intelligence levels, from narrow intelligence to general intelligence, that tries to achieve systems that think and act humanly and rationally \cite{russell2010artificial}. Current intelligent systems (including our work) are examples of narrow intelligence that use ML techniques to learn from data. Thus, we used the term ML throughout the paper.}
\end{abstract}



\received{20 February 2007}
\received[revised]{12 March 2009}
\received[accepted]{5 June 2009}

\maketitle

\section{ML Pipelines from a child-centered perspective}
The stages of a typical machine-learning pipeline can be listed in six main steps, as there's no standardization yet  \cite{trl}. Following these stages with a human-centered focus can help practitioners build natural, intuitive-to-use, high-performance intelligent systems that abide by ethical considerations. In our work, we both used pre-trained, ready-to-deploy ML models and developed new models starting from scratch to increase the affordability and potential outreach of our systems. Table \ref{table:steps} lists the steps and the corresponding questions we ask in the pipeline steps to make these projects more child-centered.

\begin{table}[]
    \centering
    \begin{tabular}{|p{.23\linewidth}|p{.77\linewidth}|}
    \hline
     ML Pipeline Stages & Child-Centered Questions  \\ 
     \hline
     Problem identification & - Is the problem focusing on data collection, model development, or interaction? \\ 
      & - Are there any child-centered use cases or guidelines for this problem? \\ 
      \hline
      Data Collection & - If there is an already existing dataset for the identified problem: How can we update it to make it more child-centered? \\
      & - If no dataset is available: What should be the steps for involving children in the data collection process? \\ 
      & - If it is a supervised problem, how can we make the process more “child-centered”? \\
      \hline
      Exploratory Analysis & - What should be the data analysis steps to check the child-centeredness of the dataset? \\
      \hline
      Model Development & - If it is a baseline architecture: Is your test cases representative enough to call the model “child-centered”? \\
      & - If it is an improved version of the previous models: Which directions do your model improve the existing models from a “child-centered” view? \\
      \hline
      Ethics considerations & - Can your data, model, or end-product interface potentially harm the privacy or well-being of children? If yes, how can you prevent it? \\
      \hline
      Interface development & - How can we involve children in making interface development more “child-centered”? What are the new challenges involving children? \\
      & - What kind of new modalities does your interface use? Can children face challenges while using these new modalities? \\
      \hline
    \end{tabular}
    \caption{Machine learning pipeline steps and our questions during these steps to make the development process more child-centered.}
    \label{table:steps}
\end{table}

\section{Case Study 1: Classroom Engagement Prediction \cite{sabuncuoglu2023multimodal, sabuncuoglu2023dashboard}}

This research aims to develop a predictive model for classroom engagement levels of students while conducting group activities. We collected data, developed baseline architectures, and created an interactive dashboard to present classroom engagement levels in a student-centric approach.

\textbf{Problem identification:} \textit{Data collection and model development- existing use cases and guidelines minimally cover this problem.} Crowded classrooms are challenging for teachers in tracking students’ engagement levels.\textit{ How can we support teachers while determining the engagement level in classroom activities?} This problem requires data from students (a video, bio-data from smartwatches, etc.) that can reveal their engagement levels and a model development that can learn from the given input. In our literature review, current vision models have the capability to understand video actions and determine engagement levels. But we needed a new dataset for training a new engagement-level classifier since the previous datasets only consider facial features in online learning. So, we deduced that we should collect a new dataset and develop a baseline architecture.

\textbf{Data Collection}  \textit{There is no dataset available. It is a supervised learning problem.} We first collected data from university students. This way, we were able to publish the dataset in an open-source format which allows the fast-pacing development of AI. Most importantly, the university students could also self-evaluate their engagement level, which allowed us to collect engagement level labels in an effective way. They also allowed us to improve our data collection activities. After following this iterative process on our data, we now aim to collect data from middle school students to fine-tune our model and develop few-shot techniques. So, we did not collect any data until we made sure every step was pedagogically appropriate for children. We aim to collect a minimum amount of data to keep the well-being and privacy of students (See Ethics Considerations below for details).

\textbf{Exploratory Data Analysis:} \textit{We aimed to learn the observable features that can contribute to learning for determining the minimum amount of data we need from children.} In this analysis, we applied dimensionality reduction, some basic classifiers, and feature engineering steps to observe the effect of observable features such as 3D face landmarks and body key points. For example, from the analysis, if we had been able to determine engagement by looking only at children’s right hands, we would have only collected video recordings of right hands to anonymize the data automatically. 

\textbf{Model Development:} \textit{Since the dataset is new, we developed baseline architectures.} In our model development, we followed transfer learning techniques to learn better using a few data. We focused on the affordability of the final setup to run the model using an already available computer and a single camera. So, we used MoViNet as our main feature extractor, which previously demonstrated success in predicting actions in streaming videos. After developing the model, we also listed the model's limitations for future practitioners. Our model only represents group activities completed with tablet computers by students with good socio-economic status.

\textbf{Ethics considerations} \textit{Bringing a camera into classrooms can reveal some concerns related to classroom surveillance.} Our data pipeline and dashboard are designed to aid students and teachers in observing engagement patterns and interpreting their engagement levels. Yet, one can be concerned about using the dashboard as a classroom surveillance tool, which can process personally-identifiable data that classifies behavior, attitudes, and preferences. Privacy concerns have been a prominent challenge in the learning analytics field. Williamson et al. list four emerging challenges while developing LADs \cite{williamson_review_2022}. For each of these emerging issues, we actively communicated with researchers, teachers, and policy-makers to make our dashboard more accessible and reliable for all.

\textbf{Interface development} \textit{We involved students in the development process by coding the interface together}. In our dashboard development, we used Observable\footnote{\url{https://observablehq.com/}} and coded it during the user studies where students commented in real-time. These interactive, modular notebooks or low/no-code platforms increased the interactivity of the sessions.

\section{Case Study 2: Tangible Programming with Everyday Materials \cite{sabuncuoglu2022kart}}

This research aims to develop an affordable programming environment to increase access to programming education. We explored the use of smartphones and tablets in group activities. In our classroom scenario, groups of students use paper programming cards and everyday objects to develop algorithms. They utilize smartphones for scanning these programs and seeing the outputs.

\textbf{Problem identification:} \textit{Using existing vision-based ML models in a pedagogically-appropriate way.} Using everyday objects as programming materials can help the abstraction capability (e.g. creating functions and variables) of tangible programming environments. However, this kind of setup requires scanning these objects using a mobile device throughout the programming activities. For example, when a student wants to represent a function that can draw a \textit{tennis ball}, the student can use a real tennis ball to represent this function. But the question is \textit{what are the system requirements to help children use these everyday materials in programming activities seamlessly?}

\textbf{Data Collection:} \textit{We did not need any new data collection to develop an object recognition system. But, our scenario required the children to collect new data throughout their programming activities.} We conducted a participatory design process where we gathered children's needs and learned about their experiences in a tangible programming setup \cite{sabuncuoglu2022exploring, sabuncuoglu2022prototyping}. In an iterative manner, we conducted three different activities that involved increased levels of programming activities (unplugged, Flappy Bird, Teachable Machine) to understand their programming object creation ability.

\textbf{Model Development:} \textit{We used a ready-to-deploy model}, MobileNetv3, from Tensorflow Hub\footnote{\url{https://tfhub.dev/}} that was already engineered to run on mobile devices. We developed an on-device data pipeline that can take a few photos in Android smartphones and use it in programming activities \footnote{\url{https://github.com/karton-project/karton-android}}.

\textbf{Ethics considerations:} \textit{No data was collected from the children}. In this system, the training and inference take place on the device. So, we do not collect any data from children. We suggest all researchers consider on-device \cite{ondevice} and federated learning \cite{fed_learning} scenarios if they plan to actively update the ML model while preserving children's privacy.

\textbf{Interface development:} \textit{Children had three roles in the development process: tester, informant, and participant \cite{druin_role_2002}}. We had the opportunity to observe children and learn their user-experience considerations, and outlook for future iterations of our software.  

\section{Conclusion}

We explained use-case \#1 in \cite{sabuncuoglu2023multimodal, sabuncuoglu2023dashboard} and use-case \#2 in \cite{sabuncuoglu2022kart} in detail. In this paper, we listed the questions that we asked regularly while collecting data, building the models and developing the interfaces throughout our ML projects. Researchers and practitioners can use these questions to check the child-centeredness of their existing projects or guide their next projects to make it more child-centered.

\section*{About the Researchers}
We are child-computer interaction researchers, building interfaces and systems \textit{for} and \textit{with} children. Our respective works have focused on affordable and accessible tangible-programming interfaces, data literacy development, computer-vision-powered classroom engagement prediction systems, AI curriculum for K12 education, and tangible early mathematics interfaces for preschool children. We have worked with various age groups (preschool to high school), parents, NGOs, and policymakers.  

\bibliographystyle{ACM-Reference-Format}
\bibliography{output}

\end{document}